\documentclass[manuscript]{aastex}

\usepackage{hyperref}

\shorttitle{Young T Dwarfs in L\,1688}
\shortauthors{Chiang et al.}

\begin{document}

\title{Discovery of Young Methane Dwarfs in the Rho Ophiuchi L\,1688 Dark Cloud}

\author{Poshih Chiang 
and W.~P. Chen 
      }
\affil{Graduate Institute of Astronomy, National Central University, 300 Jhongda Road, 
       Zhongli 32001, Taiwan}

\begin{abstract}
We report the discovery of two methane dwarfs in the dark cloud L\,1688 of the 
$\rho$ Oph star-forming region.  The two objects were among the T dwarf candidates with 
possible methane absorption and cool atmospheres, as diagnosed by infrared colors 
using deep WIRCam/CFHT {\it HK} plus {\it CH4ON} images, and {\it IRAC/Spitzer 
c2d} data.  Follow-up spectroscopic observations with the {FLAMINGOS-2/Gemini South} 
confirmed the methane absorption at 1.6~\micron.  Compared with 
spectral templates of known T dwarfs in the field, i.e., of the old populations, 
Oph\,J162738$-$245240 (Oph-T3) is a T0/T1 type, whereas Oph\,J162645$-$241949 
(Oph-T17) is consistent with a T3/T4 type in the H band but an L8/T1 in the K band.  
Compared with the BT-Settl model, both Oph-T3 and Oph-T17 are consistent with being 
cool, $\sim1000$~K and $\sim900$~K, respectively, and of low surface gravity, $\log (g)=3.5$.  
With an age no more than a couple Myr, 
these two methane dwarfs thereby represent the youngest T dwarfs ever confirmed.  
A young late L dwarf, Oph\,J162651$-$242110, was found serendipitously in our 
spectroscopic observations.
\end{abstract}


\keywords{brown dwarfs; stars: formation; infrared: stars; 
	stars: individual (Oph-L, Oph-T3, Oph-T17)
}

\section{Introduction}

Thousands of brown dwarfs have been discovered in the field by large 
surveys such as SDSS \citep{Knapp04,Chiu06,Scholz09}, 2MASS/DENIS 
\citep{Kirkpatrick00,Tinney05,Reid08,Martin10}, 
UKIDSS \citep{Burningham13}, and WISE \citep{Kirkpatrick11,Kirkpatrick12}. 
Recent discoveries of Y dwarfs push the mass of brown dwarfs down to some 10 Jupiter 
masses \citep{Kirkpatrick12}.  Furthermore, a few isolated,
or possible members of young moving groups (100--200~Myr), planetary mass objects, 
which have masses lower than the deuterium-burning limit, were recently identified 
with different spectral types, ages, and environments \citep{Delorme12,Liu13,Luhman14,
nau14,gag15}.  The current paradigm is that stars are formed out of dense molecular cloud 
cores, and planets are condensed in the protostellar disks.  The origin of substellar 
objects, however, remains unsettled.

To shed light on the issue, identification of substellar objects in a star-forming 
region, i.e., at an epoch when these objects are being formed, is a crucial first step.  
Most brown dwarfs found in nearby star-forming region so far are of late-M or L types, 
e.g. OTS\,44\ in 
Chamaeleon \citep{Oasa99,Luhman04}, 2M0437$+$2331\ in Taurus \citep{Bowler14}, and 
CFHTWIR$-$Oph33 in $\rho$ Oph \citep{Alves12}.  Toward later types, it has been a long 
battle to recognize the youngest T dwarfs to even the closest star-forming regions, 
though some {\em candidates} have been reported, e.g., in IC\,348 \citep{bur09}, and in 
Serpens \citep{spe12}, both of 1--3~Myr.  
Field brown dwarfs appear to show the transition from spectral type L (cloudy, red, and dusty) 
to T (cloudless, blue, and methane-bearing) around $\sim$1400K \citep{Kirkpatrick05}.
In contrast, young exoplanets and free-floating planetary-mass objects are expected 
to have the spectral transition at lower temperatures, possibly as the result of 
different grain physics in cool atmospheres \citep{Bowler11, Barman11, Liu13}.  
To date, three controversial cases of very young T dwarfs were claimed, 
respectively, with two in $\sigma$\,Ori, S\,Ori\,70 \citep{Zapatero02} and S\,Ori\,73 
\citep{Bihain09}, and one in $\rho$\,Oph, No.~4450 \citep{Marsh10}.  
However, recent proper motion data indicate S\,Ori\,73 likely to be a field T dwarf, 
while the nature of S\,Ori\,70 remains elusive \citep{Ramirez11}.  
On the other hand, No.\,4450 does not show prominent 
methane features.   A sample of confirmed methane dwarfs of a few million years 
old, therefore, will provide constraints to the theoretical modeling of the evolving cool 
atmospheres.  Here we report the confirmation by infrared spectroscopy of two 
T dwarfs in the L\,1688 cloud of the Rho Ophiuchi complex, 
Oph-T3 (Oph\,J162738$-$245240, RA=$16^{\rm h} 27^{\rm m} 38\fs21$, 
Decl=$-24\degr 52\arcmin 39\farcs9$, 
J2000, $H=18.38 \pm 0.06$, $H-K=0.28\pm0.09$) and 
Oph-T17 (Oph\,J162645$-$241949, RA=$16^{\rm h}26^{\rm m} 45\fs23$, 
Decl=$-24\degr 19\arcmin 49\farcs2$, 
J2000, $H=19.16\pm0.15$, $H-K=0.49 \pm 0.17$), 
following the identification numbering in Table~3 of \citet{Chiang15}.  

\section{Observations and Data Reduction}

\subsection{Photometric observations}

The T dwarfs candidates in L\,1688, the densest cloud in Rho Ophiuchi, were 
identified by cool, methane-bearing atmospheres that characterize T dwarfs \citep{Chiang15}.  
The methane absorption is diagnosed by our deep methane imaging survey 
carried out in 2010 using the {\it CH4ON} filter, centering at 1.69~\micron, 
with the {\it Wide-field InfraRed Camera (WIRCam)} attached to the 3.6-m 
{\it Canada-Frace-Hawaii Telescope (CFHT)}.  Together with the archival {\it WIRCam/CFHT} 
H-band images, the color $H-CH4ON$ serves to trace possible methane absorption near 1.6~\micron.  
In addition, the color $[3.6]-[4.5]$, available through the {\it c2d} catalog \citep{c2dcatalog}, 
is used to detect another methane feature at 3.4~\micron \citep{Patten06,Leggett10,Mace13}. 

\begin{figure}
  \begin{center}
    \includegraphics[height=0.4\textheight,angle=0]{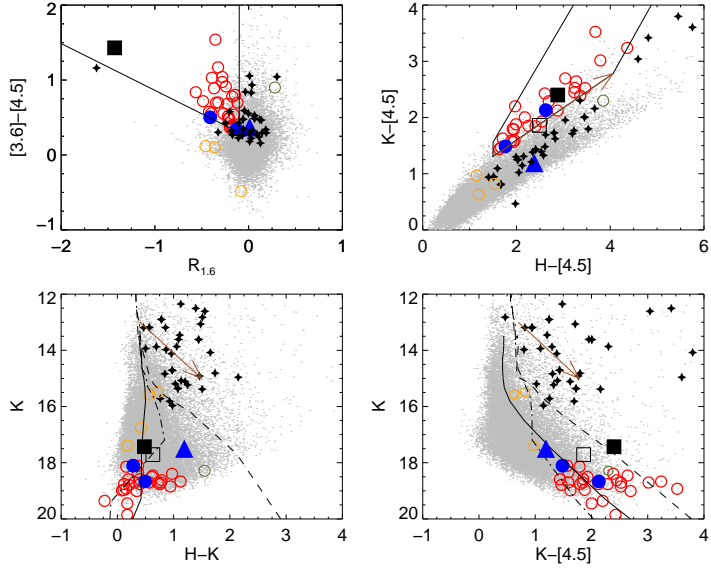}
  \end{center}
  \caption{({\it Top}) The color-color diagrams of (left) the methane indices, and (right) of the temperature indices.  
	    Gray dots represented all the sources in our data of L\,1688\ in Rho Ophiuchi.  
	    The region within which T dwarfs are identified (the ``T zone'') in each diagram is 
	    enclosed by solid lines.  Stars located in both T zones are considered T dwarf 
	    candidates (red open circles), including the two confirmed T dwarfs (blue filled circles) 
	    and one L dwarf (blue filled triangle) reported here.     
	    Asterisks represent the spectroscopically confirmed late-M and L dwarfs in 
	    $\rho$\,Oph \citep{Geers11,Alves10,Muzic12,Alves12}.  Also marked are the suspected 
	    T dwarf candidates in other star-forming regions, such as S\,Ori\,70 \citep{Zapatero02} 
	    (filled square) and No.\,4450 \citep{Marsh10} (open square).  
	    No.\,4450\ has no detected methane feature, so it is not included in the 
	    methane diagram. The T dwarf candidates in IC\,348 \citep[green open circle,][]{bur09} 
	    and in Serpens \citep[orange open circles,][]{spe12} are also marked. Note that these would not 
	    have selected by our criteria. 
	    ({\it Bottom}) The color-magnitude diagrams, with the same symbols, overlaid with the 1~Myr isochrones 
	    by the COND (solid line) \citep{Baraffe03}, DUSTY (dashed line) \citep{Chabrier00}, 
	    and BT-settl (dot-dash line) \citep{Allard12} models.  The magnitudes of S\,Ori\,70 are 
	    scaled to 130~pc from its assumed distance 352~pc of the $\sigma$\,Ori cluster.
\label{fig:1}
         }
\end{figure}

For cool temperatures, both $H-[4.5]$ and $K-[4.5]$ colors are utilized \citep{Patten06,Leggett10}.  
With these selection criteria, empirically adjusted with known M, L, and T dwarfs in the field, 
\citet{Chiang15} found a total of 28 T dwarf candidates toward L\,1688 that show evidence of both 
methane absorption and cool atmospheres.  Notwithstanding about half of these 28 candidates may 
be contaminations, likely by active galaxies or young stellar variability \citep{Chiang15}, the 
list provides a conservative, yet relatively reliable candidate sample of methane dwarfs for 
spectroscopic confirmation.  Figure~\ref{fig:1} shows the distribution of 
these candidates, including the validated T and L dwarfs presented here, in the diagnostic 
color-color and color-magnitude diagrams.

\subsection{Spectroscopic Observations}

Six T dwarf candidates were observed with {\it FLAMINGOS-2} on the {\it Gemini-South Telescope} in 2014, using 
the HK grism with a 6-pixel wide (1.08\arcsec) slit for low-resolution ($R\sim1200$) spectra.  Four of the targets 
turned out to be false positives, including one galaxy with emission lines \citep[source No.~1 in][R.A.=16:28:10.04, 
Decl.=$-24$:49:12.2 (J2000)]{Chiang15}, one background star (No.~2, R.A.=16:28:14.19, Decl.=$-24$:50:53.5, J2000),
and two low-temperature objects (No.~7, R.A.=16:27:12.80, Decl.=$-24$:49:55.2, J2000, and No.~19, 
R.A.=16:28:40.09, Decl.=$-24$:00:17.1, J2000) showing possible water absorption but no obvious methane absorption.  
In this paper, we present the results of the two confirmed methane dwarfs among the 28 candidates listed 
by \citet{Chiang15}, and the serendipitous discovery of one L dwarf.  
It took six and seven dithering points, each with a detector integration time of 300~s, in an ``ABBA'' 
pattern to complete the observations, rendering a total exposure time of 1,800 seconds for Oph-T3 and 
2,100 seconds for Oph-T17.  
An argon lamp was used for wavelength calibration.  The standard star HIP\,82271 (B9\,V) was observed to 
remove telluric absorptions.  

\begin{figure}
	\begin{center}
 \includegraphics[height=0.3\textheight, angle=0]{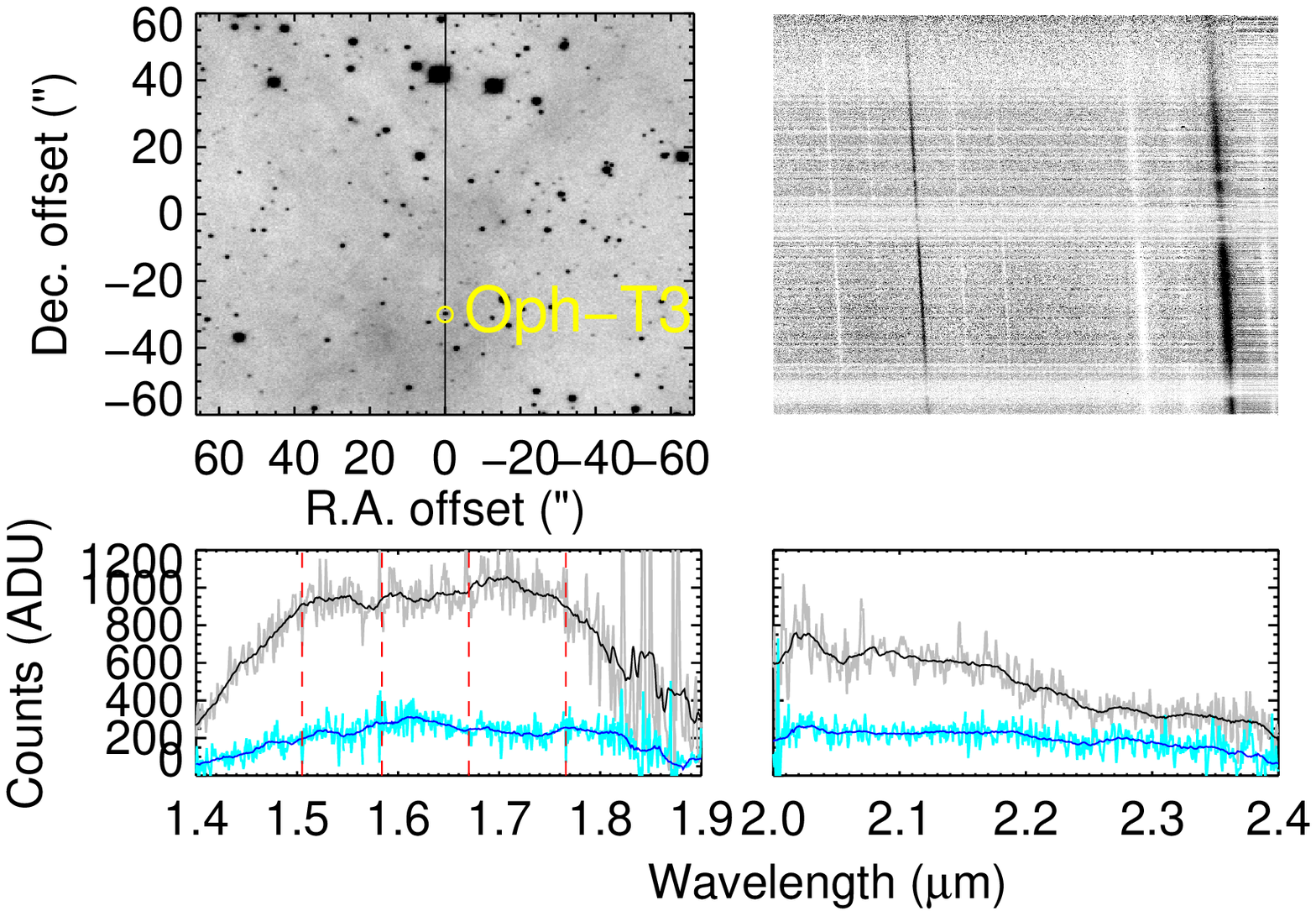}
 \includegraphics[height=0.25\textheight, angle=0]{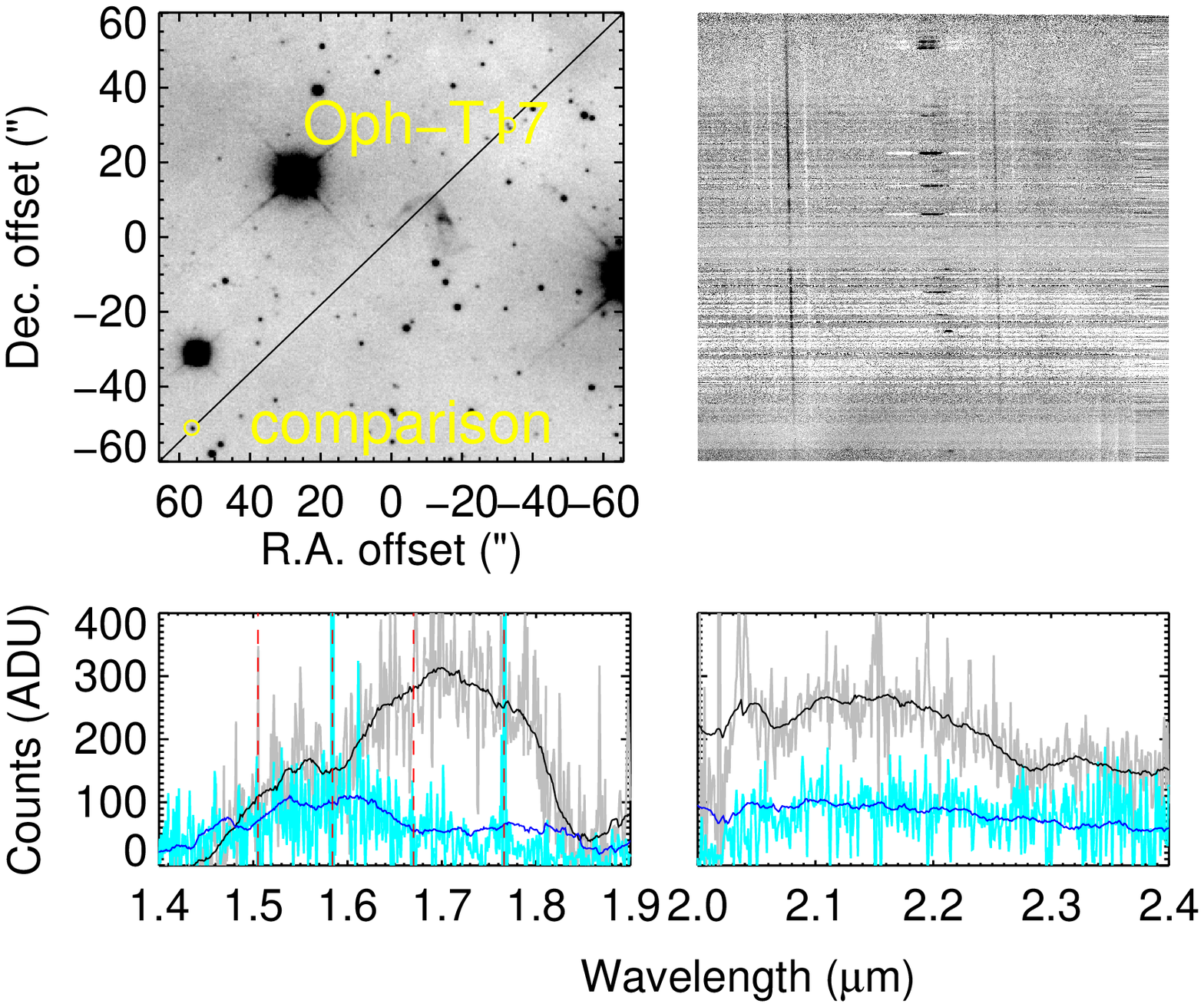}
\end{center}
 \caption{({\it Top}) The observations of Oph-T3, showing (upper-left) the WIRCam {\it CH4ON} image, 
with the target Oph-T3 around $(0,30)$, a bright star around $(0, 45)$, and 
the vertical solid line representing the position of the slit, (upper-right) the 
FLAMINGOS-2 2D dispersed image, with the thin and thick black stripes as Oph-T3 and the 
bright comparison star, (lower-left) the reduced H-band spectra for Oph-T3 (cyan) and 
the comparison (grey), each smoothed and represented by a thicker line, 
(lower-right) the reduced K-band spectra of the two stars.  The red dashed lines in the 
H band mark the prominent sky emissions.  ({\it Bottom}) The same as in the top but for Oph-T17.  
Oph-T17 is around $(-30,30)$ and its comparison around $(55,-50)$ in the {\it CH4ON} image.   In the 
dispersed image, the spectra of Oph-T17, the emission nebula, and Oph\,J162651$-$242110 are 
seen from right to left.
\label{fig:2}}
\end{figure}

Raw data were processed with the standard {\it F2/Gemini} package under 
{\it IRAF}, including flat-field correction, cutting and co-adding images, wavelength calibration, 
telluric correction, and extraction of spectra.  With the long exposures, sky emissions at 1.5, 1.58, 
1.68 and 1.77\micron\ were too bright to be properly removed so were discarded manually in 
subsequent analysis.  Oph-T17 was observed before dawn with the airmass increasing from 1.7 to 2.7, hence with 
an elevated thermal sky background.  To minimize the effect of background changes, we scaled the 
images of each dither frame to the first frame.  One additional bright star and one faint star, 
respectively, happened to be 
detected in the 2-D dispersed images of Oph-T3 and Oph-T17.    
Figure~\ref{fig:2} shows the reduced and coadded 2D dispersed images, together with the extracted 1D 
spectra of Oph-T3 and Oph-T17.  The bright star serves as a spectroscopic comparison for Oph-T3.  
The faint star coincidentally located in the slit of the Oph-T17 images, Oph\,J162651$-$242110, turned out 
to be a late L dwarf previously unrecognized.

\section{Spectral Typing}

We made two approaches for spectral classification of our candidates:~(1) by comparison
of the observed spectra with those of field brown dwarfs, and with atmospheric 
models, and (2) by spectral indices relevant to individual molecular absorption features.  

Figure~\ref{fig:3} shows the spectra of Oph-T3, Oph-T17 and Oph\,J162651$-$242110, 
all smoothed to facilitate visual comparison.  Superimposed in the figure are the 
template spectra of field L and T dwarfs taken from the SpeX Prism Spectral 
Libraries ({\url{http://pono.ucsd.edu/~adam/browndwarfs/spexprism/publications.html}), 
and the BT-Settl model spectra.   
In each case, the spectral running from the H to K band is normalized at 1.58~\micron.  

The template and model spectra have been all reddened using the empirical infrared reddening 
law prescribed by \citet{ind05}, namely 
$\log (A_{\lambda}/A_K) = 0.61 - 2.22 \log \lambda + 1.21 (\log \lambda)^2 $.
The amounts of extinction of our candidates, however, are highly uncertain.  \citet{Chiang15} 
estimated the extinction to each of their candidates on the basis of near-infrared star count.  
Oph-T17 ($A_K\sim1.8$) appears to suffer more extinction than Oph-T3 does ($A_K\sim=0.6$). 
For Oph-T17, this must be an overestimate because the object is visible, albeit fainter, at shorter 
wavelengths. On the other hand, dereddening the observed color and magnitude along the interstellar 
reddening back to the model isochrone results in $A_K\sim0.3$~mag, but this estimate is 
susceptible to the uncertain validity of the adopted model isochrone, which is itself 
something we want to check.  In Figure~\ref{fig:3}, we adopted a judicious value $A_K\sim1.2$~mag 
for Oph-T17 to illustrate the effect of extinction on the emission continuum between H and K bands. 
The extinction has little influence on the methane line diagnosis covering a relatively narrow 
wavelength range.  Likewise, $A_K\sim0.8$~mag for Oph-T3 and $A_K\sim0.3$~mag for the L dwarf 
Oph\,J162651$-$242110 are used in the plots, again for illustration, in each case by a consistent  
H to K band continuum and tracing roughly to the BT-Settl isochrone.  
A $\log (g)=3.5$ is used in the atmospheric models to compare 
to the observations.

Judging by the overall shape of the spectra, Oph-T3 should be a T0/T1 type, with the possibility 
of a late L.  For Oph-T17, the 
spectrum in the H band is best represented by a T3/T4 type, but is more consistent in the K band 
with an earlier type, L8/T1.  Note that the H-band spectrum resembles that of 
HR\,8799e \citep{Oppenheimer13}.  The spectrum of Oph\,J162651$-$242110\ in the H band exhibits 
possible water absorptions and follows reasonably well the L0 object 2M0437 in Taurus and the low-gravity 
exoplanet 2M1207b \citep{Patience10}), and in the K band is consistent with 
BT-Settl models of 1100~K and low gravity, later than an L0.     

The discrepancy in spectral typing in the H- and K-band for Oph-T17 is not surprising, given the 
spectral templates being from the old, field populations.  We compared the BT-Settl models of 
temperatures from 800 to 1100~K with various surface gravity values.  The general trend is an 
enhanced suppression of the methane feature beyond 2~\micron\ with a lowering temperature, 
i.e., with a lower peak in the 2.10--2.15~\micron\ range, relative to that near 1.6~\micron.  
Hotter than about 1000~K, the methane absorptions disappear all together.  A lower surface 
gravity tends to weaken the methane features.  There is also the possibility of composite 
spectra due to binarity \citep{mar15}.  Both Oph-T3 and Oph-T17 are consistent with being 
cool, $\sim1,000$~K, and $\sim900$~K, respectively, and low surface gravity ($\log (g)=3.5$ 
bodies.  The lack of an appreciable absorption longward to $\sim2.3$~\micron~is perhaps due to an increased 
collision-induced H$_2$ absorption with higher gravity \citep{Burgasser06}. 

Alternatively, brown dwarf spectra can be classified by spectral index, i.e., 
the ratio of the flux across a molecular absorption (e.g., by water or methane) to that of the 
continuum, as demonstrated by the sets of spectral indices published by 
\citet{McLean03, Allers13} and \citet{Mace13}.  
Because our H band data are strongly affected by sky emissions, we modified the 
wavelength ranges of each spectral index as follows: 
$CH_{4} (A) = F_{1.730}/F_{1.570} = F(1.728, 1.732) / F(1.593, 1.597)$ \citep{McLean03}, 
$(H_{2}O - H) = F_{1.470} / F_{1.600} = F(1.450, 1.490) / F(1.590, 1.610) $, and 
$(CH_{4} - H) = F_{1.648} / F_{1.600} = F(1.635, 1.660) / F(1.590, 1.610) $ \citep{Mace13}, 
where $F_{\lambda_0} = F(\lambda1, \lambda2)$ stands for the integrated flux from wavelengths 
$\lambda 1$ to $\lambda 2$, at central wavelength $\lambda_0$, all in microns.  The H$_{\rm con}$ 
index is defined by the contrast between the ``line'' and two conninuum wavelengths
\citep{Allers13}.  We adopted $\lambda_{\rm line}=1.60$, the first continuum $\lambda_1=1.46$, 
the second continuum $\lambda_2=1.69$, and the modified H$_{\rm con}$ index becomes 
$ 
  \left[ ( \lambda_{\rm line} - \lambda_1 ) \, F(1.68, 1.70) / (\lambda_2 - \lambda_1 )
	+ ( \lambda_2 - \lambda_{\rm line} ) \, F(1.45, 1.47) / (\lambda_2 - \lambda_1)  
  \right] / F(1.59, 1.61) 
$.

The spectral types of Oph-T3 and Oph-T17 can be estimated by the modified spectral indices computed for 
known T dwarfs taken from the  {\it SpeX Libraries}, as illustrated in Figure~\ref{fig:4}.  Each 
spectral index is computed for a particular molecular feature with respect to the local continuum, so 
extinction has little effect.  One sees that Oph-T3 is again consistent with a T0 type in 
all indices, whereas Oph-T17 shows a range of T0 to T4 types in different indices.  In comparison 
to the BT-Settl models, a higher temperature corresponds to a higher gravity within 800--1600~K.  
Both our targets are consistent with being cool and of low $\log (g)$, and 
Oph-T17 is cooler than Oph-T3.  
For Oph\,J162651$-$242110, the [H$_{2}$O] index suggests 
stronger water absorption than 2M0437$+$2331, a young L0 in Taurus, and the methane indices 
suggest no methane absorptions.  We therefore conclude that Oph\,J162651$-$242110 is a young 
late L-type dwarf. The spectral types thus determined are listed in Table~\ref{tab:measurements}.  

Both T dwarfs reported here are too faint in our J images.  Their (H-K) colors, 0.28~mag for 
Oph-T3 and 0.49~mag for Oph-T17, are comparable  to intermediate-age T dwarfs or planet-mass objects, 
e.g., 0.30~mag for GU\,Psc~b \citep[50--120~Myr for the AB Doradus moving group,][]{nau14}, 
0.54~mag for CFBDSIR\,2149$-$0403 \citep[also for the AB Dor moving group,][]{Delorme12}, 
and 0.60~mag for 2MASS J01033563-5515561(AB)b 
\citep[20--50~Myr for the Tucana-Horologium moving group,][]{del13}. 
Comparison of Oph-T3 and Oph-T17 does suggest a spectral sequence; in addition to 
the broad methane absorption beyond 1.6~\micron, 
both objects share similar spectral characteristics, such as the absorption between 1.52 to 
1.56~\micron.  Furthermore, Oph-T17 shows a 
more prominent methane absorption, and has a spectral running steeper than that of Oph-T3.  
We therefore conclude that OphT-17, being fainter, is cooler, i.e., of a later spectral 
type, than Oph-T3. 

Our spectroscopic confirmation substantiates the photometric selection criteria presented by 
\citet{Chiang15}.  
In the two-color diagrams of Figure~\ref{fig:1}, Oph\,J162651$-$242110 is located around the locus 
of young low-mass objects discovered in L\,1688.  In the color-magnitude diagrams, the three objects follow 
approximately the BT-settl isochrones in a spectral sequence.  
Including the L4 dwarf discovered by \citet{Alves12} and the two T dwarfs reported here, the L/T 
transition should happen between K=16 to 18~mag, as the BT-settl model suggests.



\begin{deluxetable}{ccccc}
\tablecaption{Properties of Brown Dwarfs in L\,1688\ in $\rho$ Oph}
\tablehead{
\colhead{Measurments} & \colhead{Unit} &\colhead{Oph-T3} & \colhead{Oph-T17} & \colhead{Oph\,J162651$-$242110}}
\startdata
R.A.         & hh:mm:ss.ss (J2000) & 16:27:38    & 16:26:45    & 16:26:51    \\
Dec.         & dd:mm:ss.ss (J2000)& $-$24:52:40 & $-$24:19:49 & $-$24:21:10 \\
{\it H}      & mag        & 18.38	 & 19.16	& 18.69 \\
{\it CH4ON}  & mag        & 18.51	 & 19.57	& 18.68 \\ 
{\it K$_{s}$}& mag        & 18.10	 & 18.66	& 17.50 \\         
$[3.6]$      & mag        & 16.96	 & 17.04	& 16.68 \\
$[4.5]$      & mag        & 16.62	 & 16.53	& 16.31 \\
(H$_{2}$O$-$H) & $F_{1.470}/F_{1.600}$ & 0.57 (T0)   & 0.49 (T1) & 0.34  \\
(CH$_{4}-$H)   & $F_{1.648}/F_{1.600}$ & 0.95 (T0)  & 0.70 (T3/T4) & 1.28  \\
H$_{\rm con}$ & (see text)             & 0.89 (T0)  & 0.60 (T3/T4) & 1.08 \\
H$_{2}$O (B)   & $F_{1.456}/F_{1.570}$ & 0.49 (T0)  & 0.54 (T0) & 0.23 \\
CH$_{4}$ (A)   & $F_{1.730}/F_{1.595}$ & 0.86 (T0) & 0.36  (T4/5) & 1.44  \\ 
CH$_{4}$ (B)   & $F_{2.200}/F_{2.100}$ & 0.98 (T0)  & 0.91 (T0)   & 0.93  \\ 
\enddata
\tablecomments{The wavelengths (\micron) of each modified spectral index is shown 
as the subscript in the second column.}
\label{tab:measurements}
\end{deluxetable}

\begin{figure}
\begin{center}
 \includegraphics[width=0.6\textwidth, angle=0]{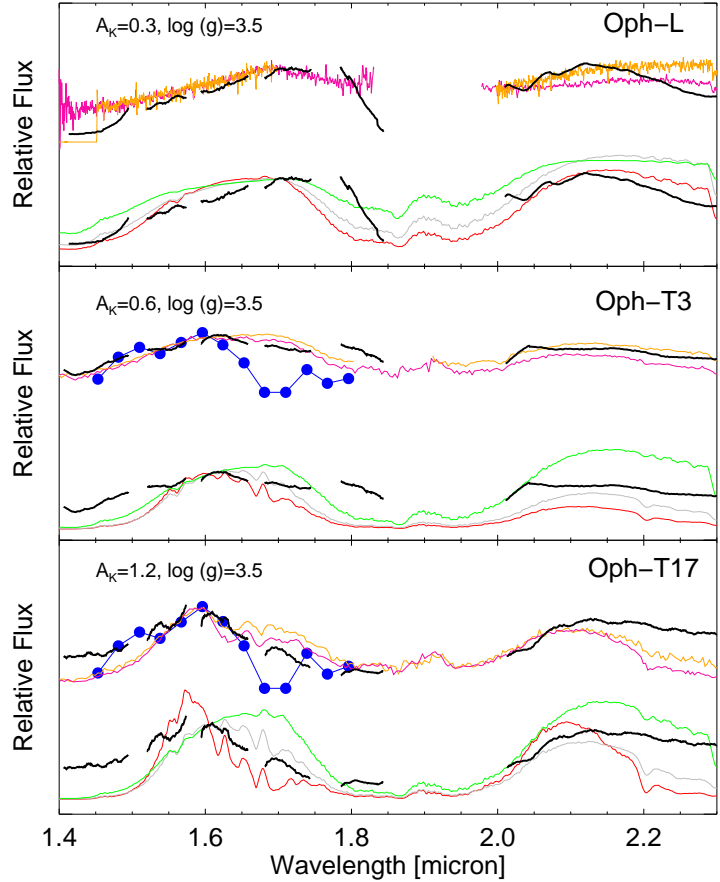}
\end{center}
 \caption{The spectra of brown dwarfs in L\,1688 of the Rho Oph region.  In every case, 
	 the top plot compares the spectrum of our target with other observations, whereas the 
	 bottom plot compares with BT-Settl models. 
	 {\it (Top)} The spectra of the young L dwarf (black), the L0 object 2M0437 (pink) 
	 in Taurus star-forming region (provided by Alves de Oliveria), the low-gravity 
	 exoplanet 2M1207b (orange) \citep{Patience10}), and the BT-Settl models of 1100~K (red), 
	 1200~K (grey), and 1300~K (green).  
	{\it (Middle)} The spectra of Oph-T3 (black), field L9 (orange) and T1 (pink) 
	templates (red), and the BT-Settl models for 900~K (red), 1000~K (grey), and 1100~K 
	(green), all for $\log g = 3.5$. The blue dot-connected line in the H band shows the data of 
	HR\,8799e \citep{Oppenheimer13}.
	{\it (Bottom)}  The spectra of Oph-T17 (black), field T3 (orange) and T4 (pink) 
	templates, and BT-Settl models for 800~K (red), 900~K (grey), and 1050~K (green), 
	all for $\log g = 3.5$. 
	The blue dot-connected line in the H band again shows HR\,8799e \citep{Oppenheimer13}.
        }
\label{fig:3}
\end{figure}

\begin{figure}
\begin{center}
 \includegraphics[height=0.9\textheight,angle=0]{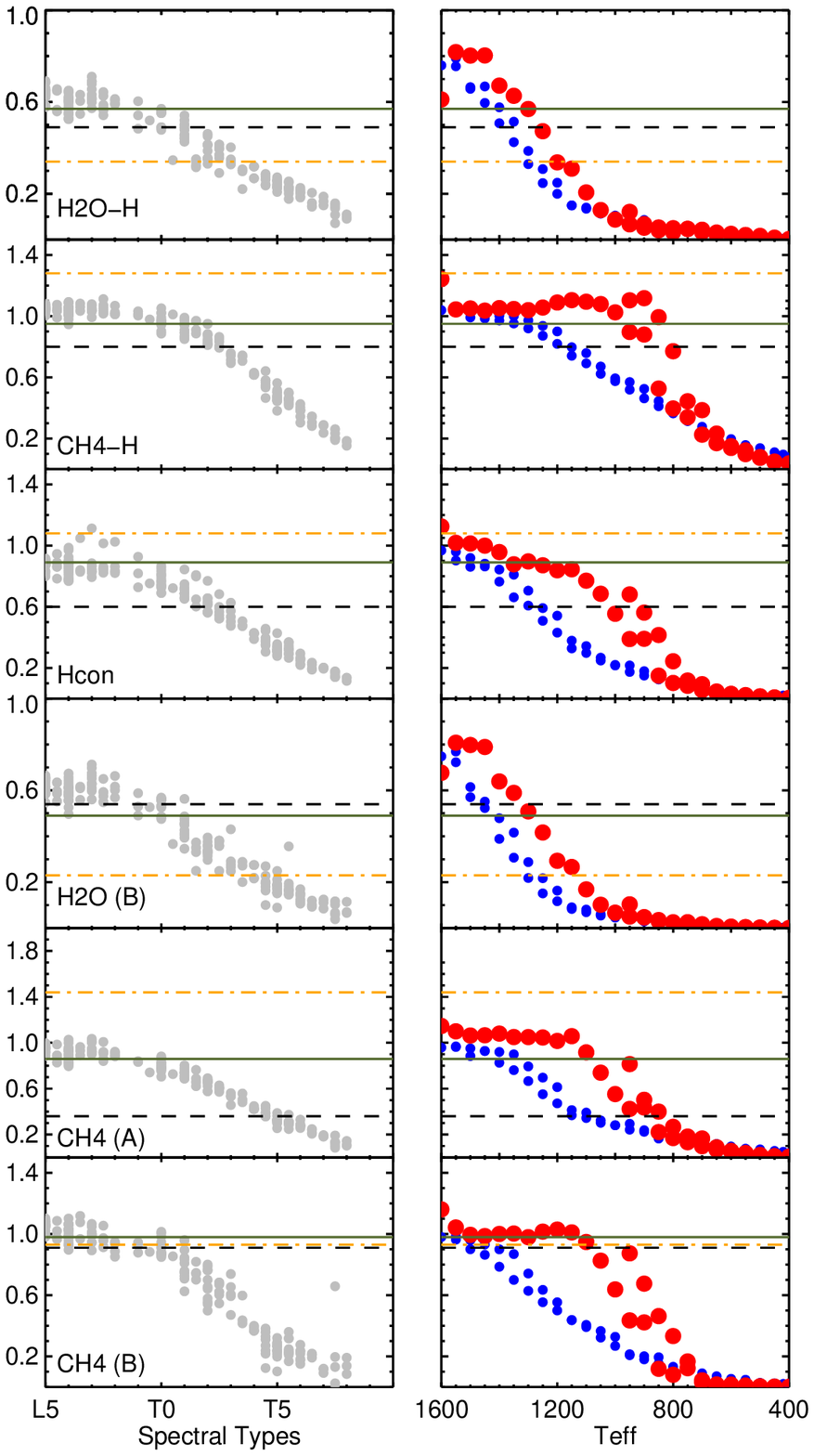}
\end{center}
\caption{Spectral indices of ({\it Left}) field brown dwarfs (grey dots), and ({\it Right}) BT-Settl models with 
	$\log (g) \ge 5.0$ (blue), and with $\log (g) \le 3.5$ (red).  
        Horizontal lines are for our Oph-T3 (solid green), Oph-T17 (dashed black), and the L object  
        (dash-dotted orange). 
        } 
  \label{fig:4}
\end{figure}

In summary, two T dwarfs and one L dwarf in L\,1688 of the $\rho$\,Oph star-forming region are identified 
on the basis of their 1.6~\micron\ methane absorptions and spectral morphology.  Their spectral types 
are determined by comparing with the spectra and spectral indices of known L and T dwarfs in the field.  
Oph\,J162738$-$245240 (Oph-T3) is of a T0 spectral type.  
Oph\,J162645$-$241949 (Oph-T17) shows a T3/T4 type in the H band but an L8/T1 type in the 
K band.  Both objects, when comparing with theoretical models, are consistent with having
cool ($\sim900$~K) and low-gravity atmospheres.  
Oph\,J162651-242110 is a late L-type object.  The three objects provide an anchor of the L/T 
transition at 1~Myr old.   The discovery of the two T dwarfs validates our identification  
method using methane and temperature sensitive colors.  Further high signal-to-noise ratio 
spectra observations obviously will determine the physical parameters much precisely.

\acknowledgments

We are grateful to Walfgang Brendner for help on data analysis, and to Catarina 
Alves de Oliveria for providing the spectrum of 2M0437$+$2331. 
We acknowledge the FLAMINGOS-2 team of Gemini South for the guidance on data reduction. 
Gratitude goes to the referee for constructive comments to substantially improve
the quality of the paper.
The project is financially supported by the MOST grant 103-2112-M-008-024-MY3.

{\it Facilities:} \facility{CFHT}, \facility{Gemini South}.

\end{document}